# Inhomogeneous Deformation of Brain Tissue During Tension Tests

Badar Rashid[a], Michel Destrade[b,a], Michael D. Gilchrist[a*]

[a]*School of Mechanical and Materials Engineering, University College Dublin, Belfield, Dublin 4, Ireland*

[b]*School of Mathematics, Statistics and Applied Mathematics, National University of Ireland Galway, Galway, Ireland*

*Corresponding Author*

Tel: + 353 1 716 1890, + 353 91 49 2344

Email: Badar.Rashid@ucdconnect.ie (B. Rashid), michel.destrade@nuigalway.ie (M. Destrade), michael.gilchrist@ucd.ie (M.D. Gilchrist)

**Abstract**

Mechanical characterization of brain tissue has been investigated extensively by various research groups over the past fifty years. These properties are particularly important for modelling Traumatic Brain Injury (TBI). In this research, we present the design and calibration of a High Rate Tension Device (HRTD) capable of performing tests up to a maximum strain rate of 90/s. We use experimental and numerical methods to investigate the effects of inhomogeneous deformation of porcine brain tissue during tension at different specimen thicknesses (4.0 – 14.0 mm), by performing tension tests at a strain rate of 30/s. One-term Ogden material parameters ($\mu$ = 4395.0 Pa, $\alpha$ = - 2.8) were derived by performing an inverse finite element analysis to model all experimental data. A similar procedure was adopted to determine Young's modulus ($E$ = 11200 Pa) of the linear elastic regime. Based on this analysis, brain specimens of aspect ratio (diameter/thickness) $S \leq 1.0$ are required to minimise the effects of inhomogeneous deformation during tension tests.

*Keywords:*   Traumatic brain injury, TBI, Impact, Compression, Shear, Ogden, Hyperelastic, Elastic

# 1 Introduction

Procedures for the mechanical characterization of soft biological tissues (such as kidney, lungs, skin and brain tissue) at quasi-static loading have been well established over the past five decades. Almost all soft tissues are now considered to be nonlinear, anisotropic and viscoelastic in nature. Determining the mechanical parameters of soft biological tissues becomes a formidable challenge at high dynamic velocities. During a severe impact to the head, brain tissue experiences a mixture of compression, tension and shear. In order to investigate the mechanisms involved in Traumatic Brain Injury (TBI), several research groups have investigated the brain's mechanical properties over a wide range of loading conditions by adopting different test protocols [1-14]. However, few tests have been performed in tension [15-17] so far.

Moreover, *diffuse axonal injury* (DAI) is the most severe form of injury, occurring at shear strains of approximately 10% – 50% and strain rates of approximately 10 – 50/s [1,3,7, 18-20]. Recently, Tamura [16] designed an apparatus to perform tests at 0.9, 4.3 and 25/s, but only the fastest of these rates is close to DAI impact speeds. The Kolsky test apparatus is usually used to perform compression tests at high strain rates, but it is more suitable for strain rates > 100/s. Based on the specific range of strain and strain rates which are injurious to axons during DAI, there is now an urgent need to develop tensile test equipment that can perform tests at strain rates up to 100 /s.

Typical tensile tests on engineering materials are performed using dog bone shaped test specimens to ensure homogeneous deformation over the required gauge length. However cylindrical specimens are more easily used for testing brain tissue because of its fragile and tacky nature, and they are usually glued at the boundaries (brain/platen interface) as an alternative to clamping. This arrangement produces an inhomogeneous deformation field near the boundaries [15]. The end effects contribute to higher magnitudes of stresses, thus resulting in steeper stress – strain curves. They also preclude the use of analytical tension – stretch relations.

Therefore, in this research, we focus on the development and calibration of a custom-designed High Rate Tension Device (HRTD) which is capable of performing tests at strain rates ≤ 90/s. In the second phase of this research, an appropriate specimen thickness and aspect ratio were determined in order to avoid any significant end effects due to inhomogeneous deformation of brain tissue during tensile tests. The inhomogeneous effects were investigated by performing several tensile tests with variable sample thicknesses of 4.0, 7.0 and 10.0 mm while maintaining a constant nominal diameter of 15.0 mm at a strain rate of 30/s. The experimental data is also analyzed numerically as a linear elastic material using Young's modulus ($E$) as well as a nonlinear hyperelastic material. This research will provide further insight into the behavior of brain tissue and the feasibility of performing reliable tension experiments on suitably sized specimens of brain tissue.

## 2  Materials and Method

### 2.1  Experimental Setup

A novel *High Rate Tension Device* (HRTD) was designed and developed [21] to perform tests at variable loading velocities (120 – 300 mm/s) to investigate inhomogeneous deformation effects on brain tissue at different specimen thicknesses, as shown in Fig.1. The major components of the apparatus include a *servo motor controlled programmable electronic actuator* (700 mm stroke, 1500 mm/s velocity, *LEFB32T-700,* SMC Pneumatics), two *± 5 N load cells* (rated output: 1.46 mV/V nominal, GSO series, Transducer Techniques) and a *Linear Variable Displacement Transducer* (range ± 25 mm, ACT1000 LVDT, RDP Electronics). The load cells were calibrated against known masses, and a multiplication factor of 13.67 N/V (determined through calibration) was used to convert voltage to load. An integrated single-supply amplifier (AD 623 G = 100, Analog Devices) with a built-in single pole low-pass filter having a cut-off frequency of 10 kHz was used. The force (N) and displacement (mm) data against time (s) were recorded for the tissue experiencing 30% strain. The images were examined to inspect that the faces of the specimens remained firmly bonded to the moving and stationary platens during extension of the brain specimen.

    A shock absorber assembly with a striker and stopper plate was developed in order to avoid any damage to the electronic assembly during the experiments, as shown in Fig. 1. The stopper plate was fixed on the machine column to stop the striker at the middle of the stroke during the uniform velocity phase, while the actuator completes its travel without producing any backward thrust (thrust absorbed by the spring) to the servo motor and other components. The actuator was run several times with and without any brain tissue specimen to ensure uniform velocity at each strain rate.

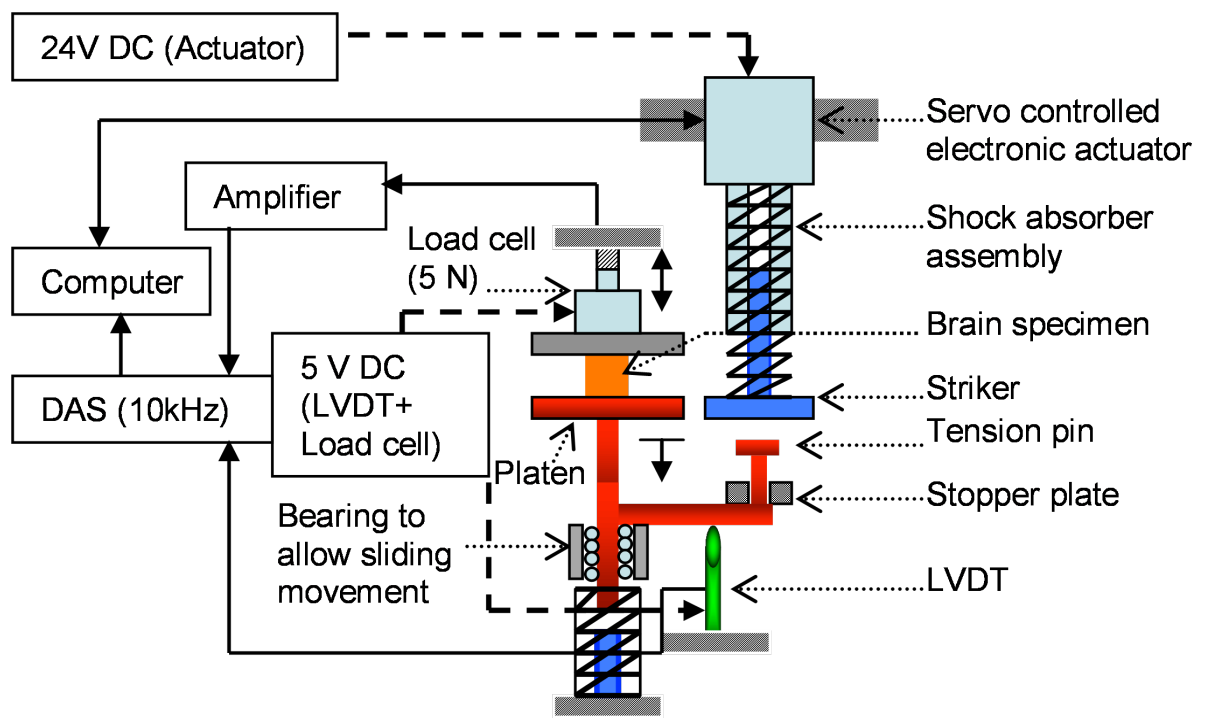

Fig. 1. Schematic diagram of the complete test apparatus. Dashed and solid lines indicate inputs and outputs respectively from the electronic components.

## 2.2 Specimen Preparation and Attachment

Ten fresh porcine brains from approximately six month old pigs were collected from a local slaughter house and tested within 3 h postmortem. Each brain was preserved in a physiological saline solution at 4 to 5°C during transportation. Then, 32 specimens were excised from 8 porcine brains (4 specimens from each brain). As shown in Fig. 2, cylindrical specimens composed of mixed white and gray matter were prepared using a circular steel die cutter (15.1 mm internal diameter) and specimens with variable thickness (4.0, 7.0 and 10.0 mm) were prepared. The time elapsed between harvesting of the first and the last specimens from each brain was 14 ~ 17 minutes. Physiological saline solution was applied to the specimens frequently during cutting and before testing in order to prevent dehydration. All samples were prepared and tested at a nominal room temperature of 22 °C and relative humidity of 34 – 35%.

The surfaces of the platens were first covered with a masking tape substrate to which a thin layer of surgical glue (Cyanoacrylate, Low-viscosity Z105880–1EA, Sigma-Aldrich) was applied. The top platen was then lowered slowly so as to just touch the top surface of the specimen. During tests, the top platen remains stationary (attached to the 5 N load cell) while the lower platen moves down to produce required tension in the specimen as shown in Fig 2. One minute settling time was sufficient to ensure proper adhesion of the specimen to the platens.

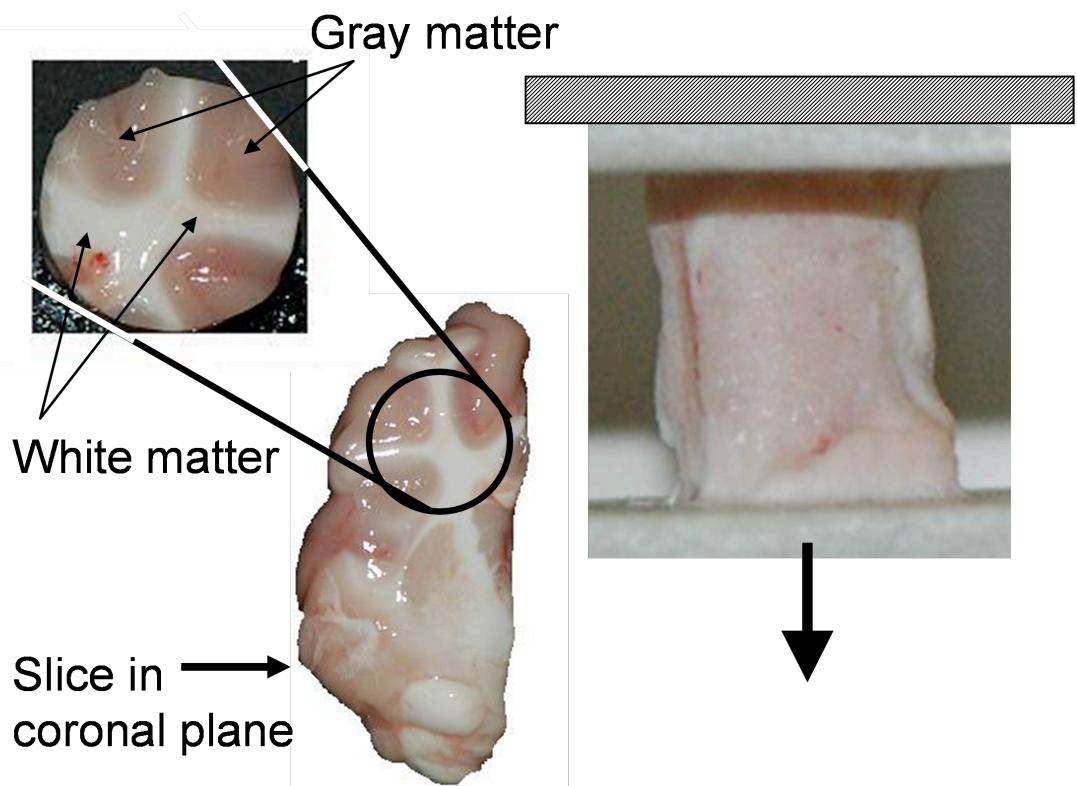

Fig. 2. Extraction of cylindrical specimens containing mixed white and gray matter (left). Specimen in stretched position after test (right).

# 3 Hyperelastic Constitutive Modelling

## 3.1 Preliminaries

In general, an isotropic hyperelastic incompressible material is characterized by a strain-energy density function $W$ which is a function of two principal strain invariants only: $W = W(I_1, I_2)$, where $I_1$ and $I_2$ are defined as [22],

$$I_1 = \lambda_1^2 + \lambda_2^2 + \lambda_3^2, \quad I_2 = \lambda_1^2\lambda_2^2 + \lambda_1^2\lambda_3^2 + \lambda_2^2\lambda_3^2. \tag{1}$$

Here $\lambda_1^2$, $\lambda_2^2$ and $\lambda_3^2$ are the squares of the principal stretch ratios, linked by the relationship, $\lambda_1\lambda_2\lambda_3 = 1$ due to incompressibility. Due to symmetry and incompressibility, the stretch ratios are of the form

$$\lambda_1 = \lambda, \quad \lambda_2 = \lambda_3 = 1/\sqrt{\lambda} \tag{2}$$

where $\lambda \geq 1$ is the stretch ratio in the direction of tension. Also, Eqs. (1) gives

$$I_1 = \lambda^2 + 2\lambda^{-1}, \quad I_2 = \lambda^{-2} + 2\lambda \tag{3}$$

so that $W$ is now a function of $\lambda$ only. The experimentally measured nominal stress was compared to the predictions of the hyperelastic models from the following relation [22], valid for homogeneous tensile tests

$$S = \frac{d\tilde{W}}{d\lambda}, \text{ where } \tilde{W}(\lambda) \equiv W(\lambda^2 + 2\lambda^{-1}, \lambda^{-2} + 2\lambda), \tag{4}$$

Soft biological tissue is often modeled well by the Ogden formulation and most of the mechanical test data available for brain tissue in the literature are fitted with an Ogden hyperelastic function [2,15,17,23,24]. The one-term Ogden hyperelastic function is given by

$$W = \frac{2\mu}{\alpha^2}\left(\lambda_1^\alpha + \lambda_2^\alpha + \lambda_3^\alpha - 3\right), \tag{5}$$

where $\mu > 0$ is the infinitesimal shear modulus, and $\alpha$ is a stiffening parameter. It yields the following nominal stress $S$, in the case of a homogeneous tensile test,

$$S = \frac{2\mu}{\alpha}\left\{\lambda^{\alpha-1} - \lambda^{-\left(\frac{\alpha}{2}+1\right)}\right\}. \tag{6}$$

# 4 Results

## 4.1 Experimentation

It was not possible to achieve homogeneous deformation conditions during tension tests due to the bonding of brain tissue (no slip conditions) at the platen/brain interfaces. This was considered to be a practical limitation of our experimental protocol. Nevertheless, an effort was made to select an appropriate specimen aspect ratio, S (diameter/thickness) which would produce negligible inhomogeneous deformation effects. Ten tensile tests were performed at each specimen thickness of 4.0 ± 0.1 mm, 7.0 ± 0.1 mm and 10.0 ± 0.1 mm at a constant strain rate of 30/s up to 30% strain while maintaining the same diameter (15.0 ± 0.1 mm). In order to maintain a constant strain rate (30/s), the required velocities were 120, 210 and 300 mm/s for the specimen thickness of 4.0, 7.0 and 10.0 mm, respectively. However, the actual measured velocities were 121 ± 1.2, 212 ± 2.0 and 306 ± 1.5 mm/s (mean ± SD). The engineering stress profiles obtained after experimentation are shown in Fig. 3. It is interesting to note that the stresses are significantly higher for the thinner specimens than for the thicker specimens at the same strain rate (30/s). It is also observed statistically, using a one-way ANOVA test, that there is significant difference (p = 0.000245) between 4.0 and 7.0 mm specimen thicknesses and similarly (p = 0.1614) between 7.0 and 10.0 mm specimen thicknesses.

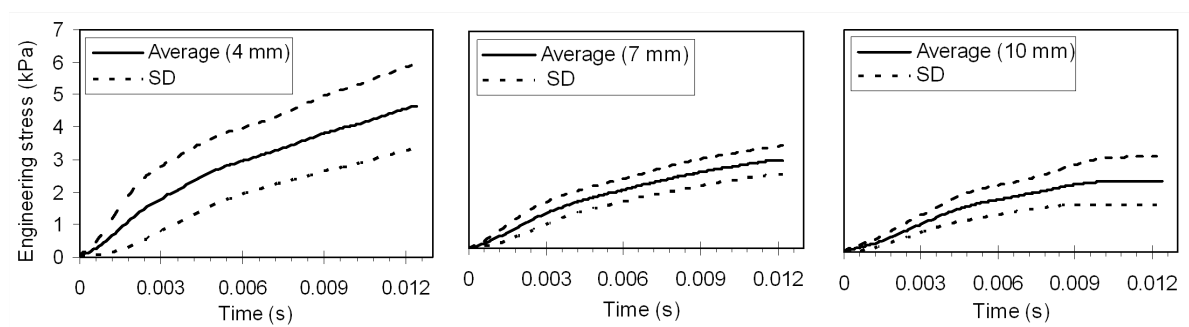

Fig. 3. Stress profiles for different thickness specimens at a constant strain rate of 30/s up to 30% strain. Note that the increasing stress with thinner specimens is a consequence of inhomogeneous deformations at the specimen – platen interfaces.

## 4.2 Finite Element Simulation

Numerical simulations were performed on different specimen thicknesses (4.0, 7.0, 10.0, 11.0, 12.0, 13.0, 14.0 mm) by applying various boundary conditions using ABAQUS 6.9/Explicit to mimic experimental conditions. One side of the cylindrical specimen was constrained in all directions whereas the other side was stretched up to 30% strain. Mass density $1040\,kg/m^3$ and C3D8R elements (hexagonal, 8-node linear brick with reduced integration) were used in the simulations. One-term Ogden material parameters ($\mu$ = 4395.0 Pa, $\alpha$ = -2.8) were obtained by performing inverse finite element analysis. The derived material parameters converged to average engineering stress (kPa) – engineering strain data obtained at different specimen thicknesses (4.0, 7.0 and 10.0 mm). This procedure was particularly important to ensure that the resulting material parameters were the same irrespective of specimen thickness. The same procedure was adopted to determine Young's modulus $E$ = 11200 Pa with a Poisson's ratio of 0.4999 as a linear elastic model, although of course the range of applicability of this model is much smaller than that of the non-linear

Ogden model. The computed modulus was approximately equal to three times $\mu$ ( $E = 3\mu$ ) as expected in incompressible elasticity. The derived elastic and hyperelastic parameters were kept constant for all the simulations performed at variable specimen thicknesses. Excellent agreement between numerically determined force (N) and engineering stress (kPa) profiles (using linear elastic and hyperelastic parameters) and experimental measurements were achieved, as shown in Fig. 4.

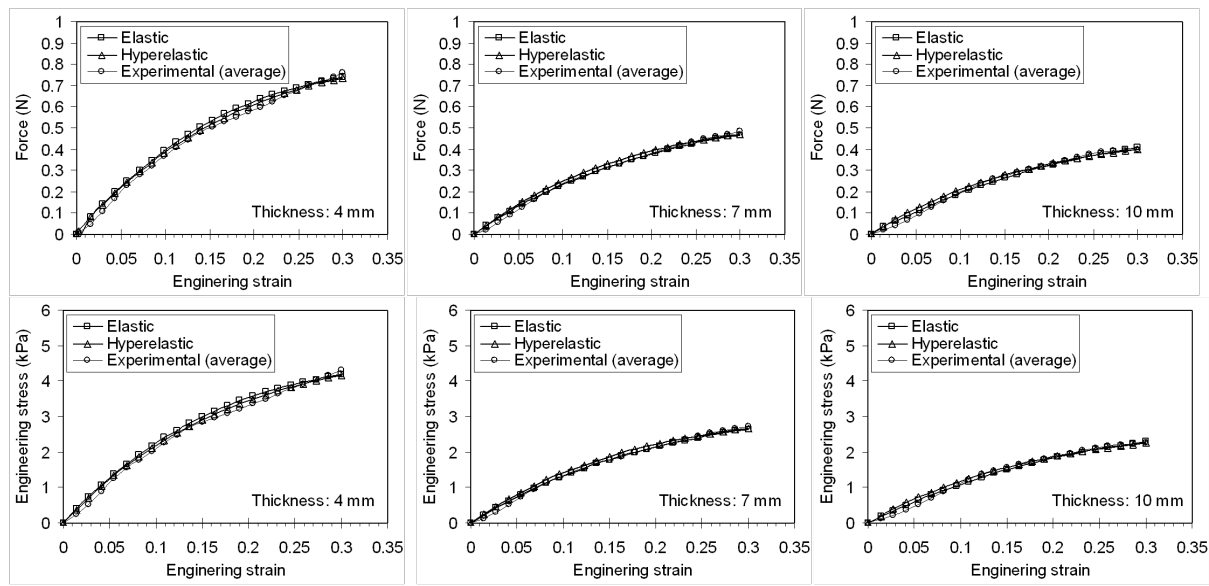

Fig. 4. Excellent agreement of elastic, hyperelastic and experimental stress (kPa) and force (N) profiles at different sample thicknesses, using Ogden material parameters ( $\mu$ = 4395.0 Pa, $\alpha$ = -2.8) and Young's modulus $E$ = 11200 Pa (at 10% strain).

It is established numerically and experimentally that the magnitudes of stresses are significantly higher with the reduction in specimen thickness or at higher aspect ratio, S (diameter/thickness). By following the same procedure, engineering stress behavior is analyzed numerically at different sample thicknesses (4 – 14.0 mm), as shown in Fig. 5 (a). Based on statistical analysis using a one-way ANOVA test, it is interesting to note that there is no significant difference (p = 0.9196) in the stress magnitudes between the thickest specimens (10.0 – 14.0 mm) i.e., at low aspect ratios (S = 1.5 – 1.07). However, there is a statistically significant difference (p = 0.000264) between specimens of 4.0 and 7.0 mm thickness and similarly (p = 0.12558) between 7.0 and 10.0 mm thick specimens. The distribution of stresses (stress S33) and strains (true strain LE) were determined numerically using hyperelastic parameters ( $\mu$ = 4395.0 Pa, $\alpha$ = -2.8) at variable specimen thicknesses. It is clearly observed that a more homogeneous stress pattern is achieved at an aspect ratio of S (diameter/thickness) $\leq$ 1.5, however stresses are significantly higher for the 4.0 mm thick specimen, as depicted in Fig. 5 (b). Similarly, the strain distributions are also analyzed at different specimen thicknesses, as shown in Fig. 5 (c). More homogeneous strain behavior is observed at aspect ratios $S \leq 1.5$ as are clearly depicted in Fig. 5 (c). The inhomogeneous strain effects at the platen ends are significantly reduced with the increase in specimen thickness. Similar stress and strain contours were also obtained when simulations were performed using a linear elastic model (Young's modulus, $E$ = 11200 Pa) and Poisson's ratio of 0.4999.

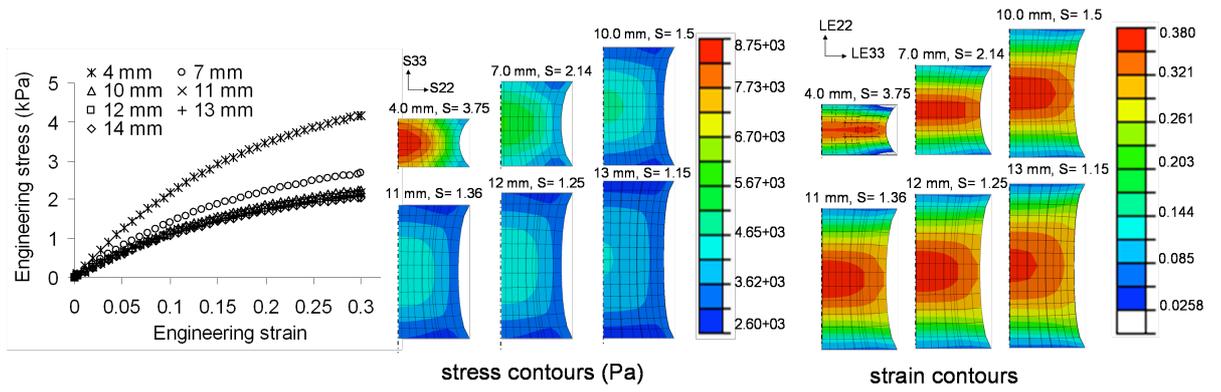

Fig. 5. Stress and strain contours at variable specimen thicknesses, at a maximum stretch ratio of 1.3.
(a) Decrease in the magnitude of engineering stress with increasing specimen thickness (4 – 14.0 mm).
(b) Significantly higher stresses are evident for specimens of 4.0 mm thickness.
(c) Homogeneous strain field is evident at 10.0 mm specimen thickness and above i.e., $S \leq 1.5$.

## 4.3 Aspect Ratio Analysis

The diameter of a test sample is also an important factor to be considered when using cylindrical specimens. Therefore, numerical simulations were performed at 10.0, 15.0 and 20.0 mm diameters for each specimen thickness (4.0, 7.0, 10.0 and 13.0 mm). Stiffening behavior is observed with larger specimen diameters; however this effect is significantly reduced at the larger specimen thickness of 13.0 mm, as shown in Fig. 6. The larger diameter produces more inhomogeneous deformation which contributes to the higher stress magnitudes. The difference between the stress profiles at aspect ratios, $S = 10/10$ and $10/13$ was also analyzed statistically using a one-way ANOVA test; the computed value of p=0.1089. The stress magnitudes are slightly higher (7%) in the case of $S = 15/10$ as compared to $S = 10/10$.

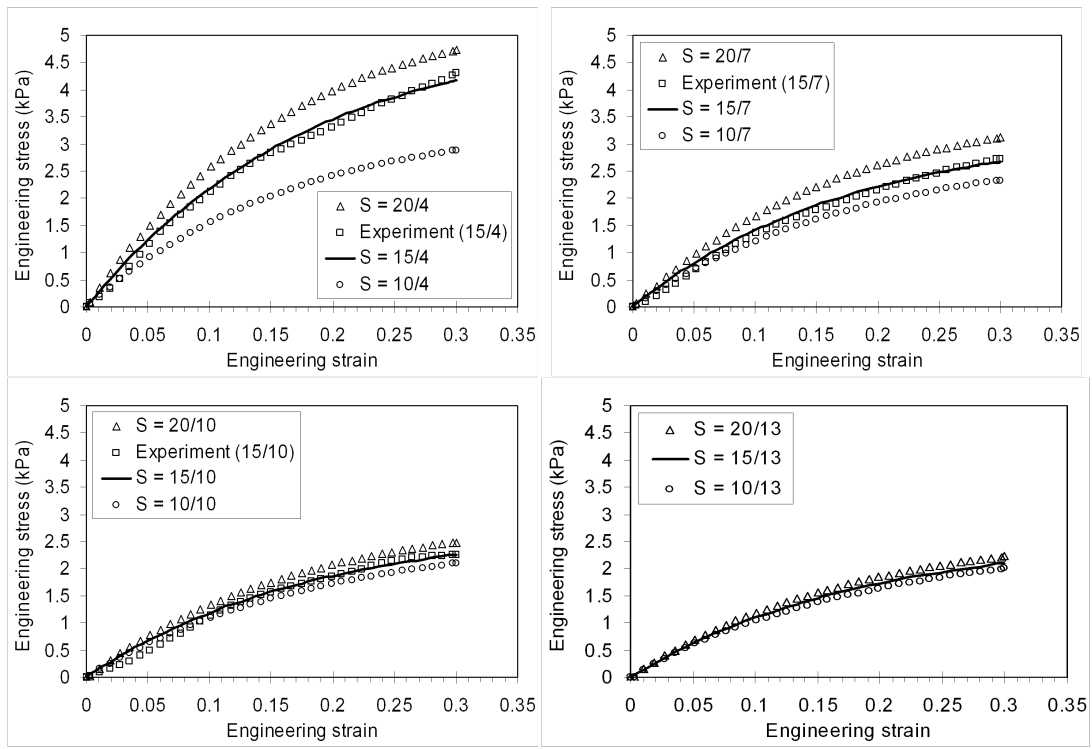

Fig. 6. Variation in engineering stress profiles at different aspect ratios, S = diameter/thickness.

# 5   Discussion

Tensile tests on cylindrical specimens of brain tissue cannot be fully characterized as classical uniaxial tension because of the specimen restriction at the boundaries (brain/platen interface). There is a strongly inhomogeneous deformation field of brain tissue near the boundaries because of its fixed attachment to the platens using surgical glue. These end effects contribute to higher magnitudes of stress, thus resulting in steeper stress – strain curves. Therefore, it was essential to determine an appropriate specimen thickness to avoid any significant end effects.

The effects of inhomogeneous deformation of brain tissue at the brain/platen interface have been analyzed experimentally and numerically using ABAQUS 6.9/Explicit. Experiments were performed using 4.0, 7.0, 10.0 mm specimen thicknesses while maintaining a constant nominal diameter of 15.0 mm. Excellent agreement was achieved between the average experimental force measured directly and the force determined numerically, as shown in Fig. 4. The engineering stress was calculated by dividing the cross-sectional area of the specimen in the reference configuration. The maximum engineering stresses at the nominal specimen thicknesses of 4.0, 7.0 and 10.0 mm were 4.5 ± 1.244 kPa, 2.73 ± 0.44 kPa, 2.24 ± 0.75 kPa (mean ± SD), respectively. The analysis was extended further by numerical simulation of specimen thicknesses from 11.0 to 14.0 mm and at variable aspect ratios, S (diameter/thickness).

Based on the present analysis, it was determined that cylindrical specimens of aspect ratio S = 10/10 or lower (10/12, 10/13) are suitable to perform tensile tests on brain tissue. Larger aspect ratio specimens do not have a sufficiently uniform stress distribution to provide meaningful results. It is noted that Miller and Chinzei [15] used cylindrical samples of diameter 30.0 mm and height 10.0 mm (S = 3) during tensile tests at quasi-static velocities (0.005, 5.0 and 500 mm/min), whereas in compression tests they used a sample height of 13 mm (S = 2.3). Tamura et al [16] on the other hand, performed tensile tests at 0.9, 4.3 and 25/s strain rates using cylindrical specimens of diameter ~ 14.0 mm and height ~ 14.0 mm (S = 1.0). This analysis suggests that the large aspect ratio samples of Miller and Chinzei [15] would have provided erroneous results due to the lack of uniform stress within their samples.

During numerical simulations up to 30% strain, it was observed that one-term Ogden hyperelastic parameters ($\mu$ = 4395.0 Pa, $\alpha$ = -2.8) and linear elasticity (Young's modulus $E$ = 11200 Pa) produced results which were in good agreement with experimental results as shown clearly in Fig. 4. The value of Young's modulus $E$ = 11200 Pa is very similar to that assumed by Morrison et al [19] ($E$ =10 kPa) in finite element simulations to predict strain fields in a stretched culture of rat brain tissue. The results of the present study at a strain rate of 30/s are slightly lower than those measured by Tamura et al [16] who performed tests at a strain rate of 25/s ($E$ = 18.6 ± 3.6 kPa).

Mesh convergence analysis was carried out by varying mesh density. The mesh is said to be converged when further mesh refinement produces a negligible change in the solution. However, in this study, the mesh was considered convergent when there was a negligible change in the numerical solution (0.8%) with further mesh refinement and the average simulation time for each specimen thickness was 60s. The total number of elements for the specimens varying in thickness from 4.0 to 14.0 mm with a constant diameter of 15.0 mm varied from 2016 to 3598, respectively. Moreover, the *accumulated artificial strain energy*, used to control *hourglass deformation* during numerical simulations, was also analyzed. It was observed that the *artificial strain energy* for the whole model as a percentage of the *total strain energy* was within the range of 1.66 – 3.4%. The significant low percentage of artificial strain energy (≤ 3.4%) observed during simulations indicates that hourglassing is not a problem.

The experimental results of this study are based on specimen dimensions of 15.0 mm diameter and variable sample thickness (4.0. 7.0 and 10.0 mm); thus there is a possibility of slightly higher stress magnitudes. Nevertheless, the primary objective of this research was to ascertain suitable specimen dimensions which should have minimum inhomogeneous deformation effects during tension tests. By following the procedure adopted in this study, the inhomogeneous deformation of the brain tissue can also be analyzed at a dynamic strain rate of 90/s in order to further understand the behavior of brain tissue at dynamic strain rates.

## 6.  Conclusions

There are four important conclusions from this work.
(i)     We demonstrated the development and calibration of a custom designed high rate tension device that is useful to obtain experimental data up to moderate strain rates of 30/s. The same experimental setup can be used to perform tests at strain rates up to 90/s.
(ii)    We have shown what dimensions of tensile test specimens of brain tissue will ensure homogeneous deformations. Cylindrical specimens of aspect ratio $S \leq 1.0$ (i.e., diameter/thickness = 10/10 or lower (10/12, 10/13)) are suitable.
(iii)   We estimated the one-term Ogden material parameters ($\mu = 4395.0$ Pa, $\alpha = -2.8$) which can be used for nonlinear hyperelastic analysis of porcine brain tissue at a strain rate ~ 30/s.
(iv)    We found Young's modulus $E = 11200$ Pa of the material, in order to analyze the behavior of brain tissue in the range of small strains.

**Acknowledgements**     We are grateful to Professor Alojz Ivankovic (UCD) for his valuable discussions. This work was supported for the first author by a Postgraduate Research Scholarship awarded in 2009 by the Irish Research Council for Science, Engineering and Technology (IRCSET), Ireland.